# Gain and Threshold Improvements of 1300 nm Lasers based on InGaAs/InAlGaAs Superlattice Active Regions

Andrey Babichev, Evgeniy Pirogov, Maksim Sobolev, Sergey Blokhin, Yuri Shernyakov, Mikhail Maximov, Andrey Lutetskiy, Nikita Pikhtin, Leonid Karachinsky, Innokenty Novikov, Anton Egorov, Si-Cong Tian, *Member, IEEE*, and Dieter Bimberg, *Life Fellow*, IEEE

*Abstract*— A detailed experimental analysis of the impact of active region design on the performance of 1300 nm lasers based on InGaAs/InAlGaAs superlattices is presented. Three different types of superlattice active regions and waveguide layer compositions were grown. Using a superlattice allows to downshift the energy position of the miniband, as compared to thin InGaAs quantum wells, having the same composition, being beneficial for high-temperature operation. Very low internal loss (~6 cm$^{-1}$), low transparency current density of ~500 A/cm$^2$, together with 46 cm$^{-1}$ modal gain and 53 % internal efficiency were observed for broad-area lasers with an active region based on a highly strained In$_{0.74}$Ga$_{0.26}$As/In$_{0.53}$Al$_{0.25}$Ga$_{0.22}$As superlattice. Characteristic temperatures $T_0$ and $T_1$ were improved up to 76 K and 100 K, respectively. These data suggest that such superlattices have also the potential to much improve VCSEL properties at this wavelength.

## I. INTRODUCTION

Short-wavelength infrared (SWIR) vertical-cavity surface-emitting lasers (VCSELs) are essential e.g. for all kinds of sensing applications due to their eye-safety [1].

Trumpf Photonic Components recently announced plans to mass produce InP-based VCSELs in the 1300 to 2000 nm range [1]. TriEye Ltd and Vertilas GmbH have developed 1300 nm SWIR sensors based on hybrid InP-based VCSELs [2].

As an alternative, IQE PLC has presented GaAs-based VCSELs using diluted nitride hybrid epi-growth technology. The high level of Shockley-Read-Hall (SRH)-recombination, the difficulty to control low nitrogen-composition layers [3], and the low reliability of metalorganic vapor-phase epitaxy (MOVPE) for the growth of diluted nitrides leads to using plasma assisted molecular-beam epitaxy (MBE) for growing the active region [3]. A 1380 nm VCSEL platform based on 200 mm diameter GaAs or Ge substrates has been demonstrated for sensing applications [4]. Typical output power was more than 1 mW, and reliability exceeded 1000 hours in CW mode [3].

High-speed SWIR VCSELs are essential for intra-data-center connections beyond ~100–200 meters. Recently, 40 Gbps non return to zero and 53 Gbaud 4-level Pulse Amplitude Modulation for single-mode SWIR VCSELs has been demonstrated by Vertilas [5–7]. An InP-based short-cavity design along with two high-contrast dielectric mirrors, mounted on a pseudosubstrate electroplated by gold provides high output power of 4 mW [7, 8] and modulation bandwidth of up to 17 GHz [7–9].

For 3D sensing applications, Vertilas developed a robust technology based on VCSELs with single semiconductor mirrors [10, 11]. VCSELs with top semiconductor mirrors allows the implementation of versatile laser arrays that are monolithically integrated (laterally connected). VCSELs based on this hybrid technology have passed 25000 hours of accelerating aging and meet Telcordia compliant qualification [12]. About 8 W quasi CW output power was realized for an 800 VCSELs array [12]. Due to the use of the single semiconductor mirror, the maximum bandwidth of the VCSEL with long-cavity design was limited to 10 GHz [10, 11, 13] and the maximum output power was about 4 mW [11].

Similar parameters (11.5 GHz bandwidths [14, 15] and 6 mW output power) have been demonstrated for wafer-fused (WF) 1300 nm VCSELs grown by MOVPE.

WF VCSELs grown by MOVPE [15, 16] as well as hybrid VCSELs with active region grown by MBE [17, 18], use

Manuscript received 02 August 2024; revised XX XXXX 2024; accepted XX XXXX 2024. Date of publication XX XXXX 2024; date of current version XX XXXX 2024. The results were obtained within a Russian-Chinese project, with the financial support of the Russian side by the Ministry of Science and Higher Education of the Russian Federation (grant agreement in the form of a subsidy from the federal budget No. 075-15-2023-579 dated August 11, 2023) and with the financial support of the Chinese side by the National Key R&D Program of China (2023YFE0111200). *(Corresponding authors: Andrey Babichev, Si-Cong Tian and Dieter Bimberg)*

Andrey Babichev, Leonid Karachinsky, Innokenty Novikov, and Anton Egorov are with ITMO University, 197101 Saint Petersburg, Russia (e-mail: a.babichev@itmo.ru);

Evgeniy Pirogov, Maksim Sobolev, and Mikhail Maximov are with Alferov University, 194021 Saint Petersburg, Russia;

Sergey Blokhin, Yuri Shernyakov, Mikhail Maximov, Andrey Lutetskiy, and Nikita Pikhtin are with Ioffe Institute, 194021 Saint Petersburg, Russia;

Si-Cong Tian, and Dieter Bimberg are with the Bimberg Chinese-German Center for Green Photonics, Changchun Institute of Optics, Fine Mechanics and Physics (CIOMP), Chinese Academy of Sciences (CAS), Changchun 130033, China, and also with the Center of Nanophotonics, Institute of Solid State Physics, Technische Universität Berlin, 10623 Berlin, Germany (e-mail: tiansicong@ciomp.ac.cn, bimberg@physik.tu-berlin.de).

TABLE I
THE THRESHOLD MODAL GAIN, TRANSPARENCY CURRENT DENSITY, INTERNAL EFFICIENCY AND INTERNAL LOSS

| Sample No. | Type of active region | Threshold modal gain ($\Gamma \cdot G_0$), cm$^{-1}$ | $J_{tr}$, A/cm$^2$ | $\eta_i$, % | $\alpha_i$, cm$^{-1}$ |
|---|---|---|---|---|---|
| 1 | 21 periods of In$_{0.60}$Ga$_{0.4}$As (1.3 nm thick)/In$_{0.53}$Al$_{0.25}$Ga$_{0.22}$As (2 nm thick) | 45 | 590 | 54 | 8 |
| 2 | 23 periods of In$_{0.74}$Ga$_{0.26}$As (1.0 nm thick)/In$_{0.53}$Al$_{0.25}$Ga$_{0.22}$As (2 nm thick) | 46 | 500 | 53 | 6 |
| 3 | 27 periods of In$_{0.74}$Ga$_{0.26}$As (0.6 nm thick)/In$_{0.53}$Al$_{0.20}$Ga$_{0.27}$As (2 nm thick) | 49 | 640 | 43 | 10 |

compressively strained InAlGaAs quantum wells (QWs) to increase the modulation bandwidth. The large Al-content in strained InAlGaAs QWs (~ 18%) yields in weaker carrier confinement [19] and increase the SRH-recombination and threshold currents [16, 17].

Previously, the use of InGaAs QWs has shown its effectiveness (minimization of the laser threshold) as the active region of 1550 nm VCSELs based on hybrid [20–26] or WF [27–30] technologies. Using narrow InGaAs QWs for the 1300 nm spectral range looks inefficient due to the significant thermal escape [31], which cannot be compensated by increasing the number of QWs, as noted earlier for 1550 nm VCSELs [2, 28–30].

InGaAs-InP superlattices (SLs) present an alternative to InGaAs-InP QWs. SLs allow to downshift the bottom of the miniband [32–34] as compared to the ground state of uncoupled QWs, improving the temperature stability of the laser. The optical confinement factor of SLs is increased by a factor of 2 compared to QWs case. Thus, the modal gain is increasing and the SRH-recombination is reduced, decreasing the internal loss for 1300 nm VCSELs [35].

In this paper we study the gain and threshold of InGaAs-InP superlattice-based broad-area (BA) lasers as a case study essential for 1300 nm VCSELs development. The effect of strain on laser performance and temperature dependence, showing a road to significant improvements of VCSELs-based on SLs.

## II. SWIR VCSELs ACTIVE REGION DESIGN

SWIR VSCELs have used strain-compensated active regions based on InAlGaAs QWs [15, 16, 36–38]. As a result, due to partial strain compensation in the tensile barrier layers, it is possible to realize good crystallographic properties for a strain mismatch up to 1.6 % [15]. Previously, we developed high power WF VCSELs based on In$_{0.6}$Ga$_{0.4}$As (0.68 eV)/In$_{0.53}$Al$_{0.20}$Ga$_{0.27}$As (1.00 eV) forming a SL [35, 39, 40], with a strain mismatch $\varepsilon$ between InGaAs and InP of about 0.48 %. Barriers were nearly lattice matched to InP ($\varepsilon$ = 0.02%).

The first structure studied here also uses a superlattice with a similar strain mismatch between InGaAs and InP ($\varepsilon$ =0.48 %). The bandgap of barrier layers was however increased to 1.07 eV to improve the temperature dependent properties, like threshold current density. Structure No. 2 shows a large strain mismatch ($\varepsilon$=1.44%) between In$_{0.74}$Ga$_{0.26}$As (0.53 eV) and InP. This structure is intended to study the strain effect on the material gain, as well as on the transparency carrier density. The third structure is distinguished from structure №2 by the barrier composition (In$_{0.53}$Al$_{0.20}$Ga$_{0.27}$As). The details of structures are presented in Table 1.

## III. STRUCTURE AND TEST

All structures were grown by MBE on a Riber 49 mass production system. A n-doped emitter layer being 200 nm thick with a doping level of $2\times10^{17}$ cm$^{-3}$ was grown on an n-doped InP substrate with carrier concentration of $(1–3)\times10^{18}$ cm$^{-3}$. The confinement layers consist of In$_{0.52}$Al$_{0.48}$As. The waveguide claddings were formed from 1 µm thick InAlGaAs layers of the same composition as the active region barriers. A single-step separate confinement hetero (SCH) structure was formed for comparative analysis of different active region designs. Short-period InGaAs/InAlGaAs superlattices were used as active regions for each structure. The thickness of the active region was close to 70 nm, which coincided with the width of the optical field antinode in a VCSEL [35]. The p-emitter was formed by three 500 nm thick In$_{0.52}$Al$_{0.48}$As layers with doping level $1\times10^{17}$ cm$^{-3}$, $5\times10^{17}$ cm$^{-3}$ and $1\times10^{18}$ cm$^{-3}$, respectively. In$_{0.53}$Ga$_{0.47}$As layers with a total thickness of 150 nm and a carrier concentration of $1\times10^{19}$ cm$^{-3}$ were used as contact layers.

The peak of the low excitation power photoluminescence (PL) spectrum at room temperature was at 1286 nm, 1289 nm, and 1285 nm for structure No. 1, 2 and 3, the full width at half maximum of the PL spectrum was ~ 44 meV, 47 meV, and 44 meV for structures No. 1, 2 and 3, respectively.

BA edge-emitting lasers with different cavity length have been fabricated. Light-current-voltage (LIV) characteristics and multi-mode lasing spectra were measured and analyzed. The total width and current pumping width of the BA lasers were 400 µm and 100 µm, respectively. Outside the pumping regions, the InGaAs contact layer was etched into the underlying p-emitter layer and SiO$_2$ was deposited using magnetron sputtering, followed by opening a 100-µm-wide window in the contact region [41]. After substrate lapping, a bottom contact was formed, followed by cleaving of the facets.

LIV measurements were carried out in a pulse mode (300 ns pulse duration and 4 kHz frequency) in the temperature range of 20°C– 80°C. Threshold current and external quantum efficiency were measured using a calibrated Ge large-area (10 mm diameter) photodiode (J16-P1-R10M-HS model from Teledyne Technologies Inc.).

## IV. LASER CHARACTERIZATION

### A. Static characteristics measured at 20°C

The LIV dependences of BA lasers with different cavity lengths were measured at 20°C for all three different designs. Figure 1, top panel plots the inverse external quantum efficiency as a function of the inverse mirror loss ($1/\alpha_m$). The external quantum efficiency is defined as $\eta_{ext}=2\eta_s/h\nu$, where $h\nu$ is the photon energy (cf. Figure 2, inset) and $\eta_s$ is the slope efficiency determined from the front facet.

The maximum external quantum efficiency for 1 mm long lasers is found to be 37% for the second structure exceeding the largest $\eta_{ext}$ values of 31 % reported so far for InAlGaAs–InP SCH BA lasers [18].

The internal quantum efficiency $\eta_i$, as well as the internal loss $\alpha_i$ were determined by a linear approximation using the formula $1/\eta_{ext} = (\alpha_i/\alpha_m+1)/\eta_i$. The mirror loss were calculated based on the facet reflection coefficient of 0.32. The results are again summarized in Table 1.

The internal loss values were in the range of 6–10 cm$^{-1}$ for the different laser designs. The lasers based on structures No. 2 demonstrate the lowest $\alpha_i$ value (6 cm$^{-1}$), which is 25% less than the minimum internal loss (~ 8 cm$^{-1}$) for lasers based on compressively strained InAlGaAs QWs ($\varepsilon$ = 1.4%) [18].

The lasers fabricated from structures No. 1 and No. 2 demonstrated an internal quantum efficiency close to 53% (cf. Table 1), which is similar to that reported for InAlGaAs–InP SCH lasers with the same strain mismatch ($\varepsilon$ = 1.4%) [18].

The threshold current density determined for lasers with different cavity length was used to evaluate the gain. Taking into account gain saturation at large carrier densities, the dependence of material gain on current density can be determined by the expression presented in Ref. 42, 43 and adopted for SL-based active regions [35, 44]: $g = G_0 \cdot (\ln(\eta_i \cdot j/j_{tr}))$, where $j_{tr}$ is the transparency current density and $G_0$ is the gain coefficient. As a result, the threshold current density can be expressed as follows: $j_{th} = j_{tr}/\eta_i \cdot \exp((\alpha_i+\alpha_m)/\Gamma \cdot G_0)$, where $\Gamma$ denotes the optical confinement factor for the SL. Using a semi-logarithmic approximation of the threshold current density (linear approximation of the dependence of $\ln(j_{th})$ on $\alpha_m$), one can estimate the transparency current density as well as threshold modal gain ($\Gamma \cdot G_0$ value). The fitting results are summarized in Table 1 and presented in Figure 1, bottom.

For lasers based on the first structure, the transparency carrier density was about 590 A/cm$^2$, and for lasers based on the structure No. 2, the $j_{tr}$ value was 500 A/cm$^2$ due to the strain effect [45]. For structures with a lower bandgap of the barriers, the increase in transparent current density is more pronounced. As a result, lasers based on the third structure exhibit an increased transparency carrier density (640 A/cm$^2$). Previously, a still larger $j_{tr}$ value (680 A/cm$^2$) was observed for InAlGaAs–InP SCH BA lasers [18].

After the analysis of the transparency carrier density, we will discuss the gain variation. For lasers based on the first and second structures, the modal gain is similar (~ 45 cm$^{-1}$). Structure No. 3 exhibits the largest $\Gamma \cdot G_0$ value (49 cm$^{-1}$). This modal gain exceeds that for InAlGaAs–InP SCH lasers with similar ($\varepsilon$ = 1.4%) strain mismatch (43 cm$^{-1}$ [18]).

To analyze the modal gain results, it is necessary to keep in mind that the strain mismatch and thickness of InGaAs were changed simultaneously in order to maintain a constant emission wavelength (cf. Table I). As a result, both effects can influence material gain.

In general, a decrease in the QWs width leads to an increase in the distance between energy subbands and shifts the levels position towards the barrier edges. As a result, the band edge occupation is typical increased for thin QWs and can enlarge the gain [33]. A decrease of the overlap between electron and hole subband envelope functions compensates the first effect. Thus, the material gain of thin QWs is less than that in thick QWs [33]. This effect is reduced for SL active regions, in which, due to the formation of a $Mb$, a significant overlap of electron and hole wavefunctions is observed. Moreover, the coupling between the QWs yields an increase in the band edge density of states and lowers the quasi-Fermi level position [33].

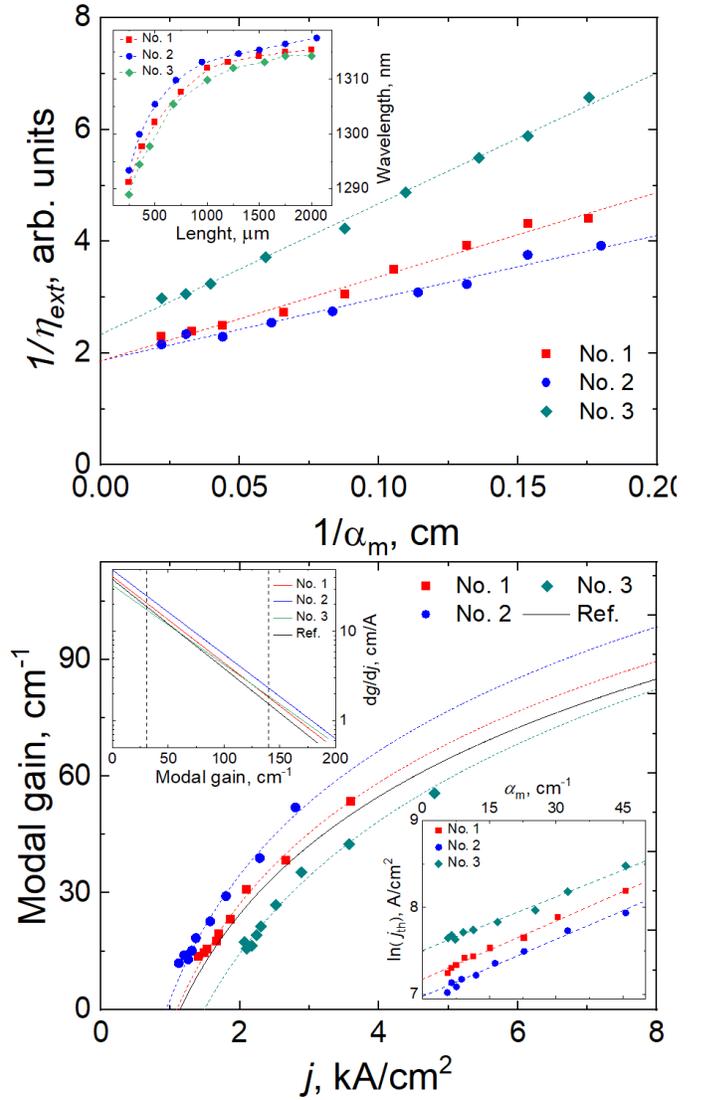

**Fig. 1.** Inverse external quantum efficiency as a function of inverse mirror loss (top panel). The inset of top panel shows the lasing wavelengths for lasers with different cavity length; Threshold modal gain as a function of current density (bottom panel). The solid line shows the modal gain calculated for the structure studied in Ref. 18. The top inset demonstrates the $dg/dj$ as a function of modal gain. The bottom inset shows the $\ln(j_{th})$ ($\alpha_m$) dependences.

An increase in the compressive strain mismatch between QWs and substrate affects the quasi-Fermi levels position by two reasons. The first is associated with a decrease of the valence Fermi level and an increase of hole occupation (increasing the material gain) [33]. The second effect is an increase of the spacing between holes subbands with compressive strain [45]. The enlarge of the barrier energy also affects the subband energy spacing. All these effects reduce the valence band edge density of states. As a result, a reduction in the transparency carrier density and an increase in the gain coefficient $G_0$ can be observed [45].

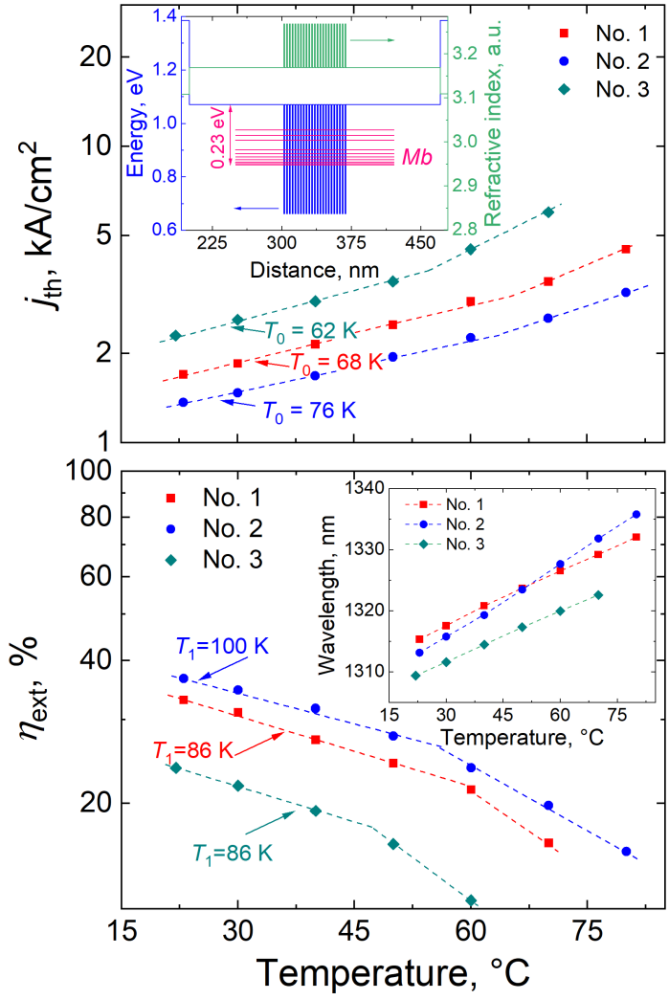

**Fig. 2.** Temperature dependence of the threshold current density (top panel) and the total external quantum efficiency (bottom panel) for lasers based on three structures. Each panel is in semi logarithmic scale. The inset at the top panel shows the conduction band energy diagram of the second structure. The inset at the bottom panel depicts the lasing wavelengths at different temperatures.

The strain mismatch between barriers and InP also affects the gain value. The use of tensile strained barriers [15, 16, 36–38] allows increasing the number of QWs in the strain-compensated active region, but significantly deteriorates the material gain (to about 2.5 times for the case of $In_{0.63}Ga_{0.37}As/In_{0.43}Al_{0.17}Ga_{0.40}As$ QWs compared to the structure with unstrained barriers) due to the upward shift of the light hole subbands [46]. In contrast, the use of compressively strained barriers increases the material gain of the active region based on compressively strained QWs.

We calculated our hole level positions using by following the approach in Ref. 47. The transition from the first to the second design leads to an increase in the subband spacing to about 23 meV. The third design exhibits a subband spacing that is being approximately 8 meV less than that of the first design. These results agree with the experimentally obtained changes in the transparency current density (cf. Table I).

In general, the use of narrow coupled QWs will result in a decrease in differential gain compared to uncoupled QWs [33, 48] due to the $Mb$ formation. At the same time, differential gain is inversely proportional to the QWs number [49]. To clarify this issue, an estimate of the differential gain from experiment is required. To compare the gain characteristics, the modal gain was determined as a function of current density (cf. Figure 1, bottom panel). The differential gain can be estimated from the slope of the $g(j)$ dependence (cf. the top inset of Figure 1, bottom panel). The maximal differential gain is found for lasers based on the second structure. Figure 1 also shows results for InAlGaAs–InP SCH lasers based on four thick QWs ($\varepsilon = 1.4\%$) [18]. For modal gain less than 60 cm$^{-1}$, the minimum d$g$/d$j$ value corresponds to structure No. 3. When this $g$ value is exceeded, the minimum d$g$/d$j$ value is demonstrated by a reference structure based on InAlGaAs QWs [18].

To quantify the differential gain, we estimate the modal gain $G_0$ for a previously studied SL-based VCSEL [35], which is about 53 cm$^{-1}$ at 20 °C (6 μm buried tunnel junction case). This value was calculated taking into account $G_0$ at 60 °C (~30 cm$^{-1}$), d$G_0$/d$T$ [50] and the gain to cavity detuning value (~20 nm at 20°C). The position of the modal gain at threshold and roll-over current is marked by dashed lines on the inset of Figure 1 and equaled to 30 and 140 cm$^{-1}$, respectively. At these values of modal gain, the difference between d$g$/d$j$ values of the second design and InAlGaAs–InP QWs [18] is about 33 % and 53% close to threshold and roll-over current, respectively.

In summary, it can be concluded that the differential gain for the SL-based active regions is significantly increased compared to InAlGaAs QWs at the same strain mismatch.

*B. Static characteristics measured at different temperatures*

The temperature characteristics of the threshold current density, as well as external quantum efficiency determine the ability of the lasers to operate at high temperatures. The characteristic temperature $T_0$ is derived using the expression $1/T_0 = d\ln(j_{th})/dT$ [51].

The temperature dependences of the threshold current density for lasers with 1 mm length are presented on Figure 2 for all three structures. The $j_{th}(T)$ dependence can be divided into two regions (cf. Figure 2, top panel). Below a certain temperature, called critical ($T_c$) [52, 53], Auger recombination significantly affects $j_{th}(T)$ [54]. A change in the slope of the $j_{th}(T)$ was observed above 65 °C for lasers based on the first and second structures. The $j_{th}(T)$ dependence for lasers based on the third structure demonstrates a clear change in slope at 55 °C.

As one can see (cf. Figure 2, top panel), the characteristic temperatures $T_0$ for our structures are in the range of 62–76 K. A minimum $T_0$ value (~62 K) was found for lasers based on the third structure between 20 °C and 55 °C. Lasers based on the first and second structures have a characteristic temperature $T_0$ of 68 K and 76 K in the range from 20 °C and 65 °C.

To compare the characteristic temperature $T_0$, the position of $Mb$ can be estimated [55]. The first and second structures, in which the InGaAs thickness was increased, compared to the active region of VCSELs studied in Ref. 35, show a significant downshift of the $Mb$ bottom edge from -170 meV [35] to -226 meV and -225 meV, respectively, below the barrier edge. The third structure exhibits a $Mb$ position being 167 meV below the barrier edge. As a result, a decrease in the InGaAs thickness (structure No. 3) makes it possible to increase the differential gain hovewer being accompanied by low temperature stability.

The largest characteristic temperatures $T_0$ for BA lasers based on thick InAlGaAs QWs ($\varepsilon$ = 1.4%) is 81 K [18]. To comparative analysis of the characteristic temperature $T_0$, it is possible to estimate the position of energy levels for active region in Ref. 18 and for studied second structure. These two active regions design have the same thickness (~70 nm) and similar modal gain (~45 cm$^{-1}$).

Four $In_{0.73}Al_{0.16}Ga_{0.11}As$ QWs separated by $In_{0.52}Al_{0.32}Ga_{0.16}As$ barriers [18] exhibit two electron levels, located about -45 meV and -270 meV below the barrier edge.

Simulations show that the $Mb$ band position of the SL is downshifted by about 80 meV compared to thin uncoupled $In_{0.74}Ga_{0.26}As/In_{0.53}Al_{0.25}Ga_{0.22}As$ QWs. The $Mb$ bottom edge is about -225 meV from the barrier edge (cf. Figure 2, inset). And a 56 meV wide region is formed by crossed ground states of coupled QWs. In addition, three excited states of coupled QWs are observed at -133 meV, -115 meV and -95 meV from the barrier edge. As a result, the bottom edge of the $Mb$ is located ~ 45 meV above the ground level in thick $In_{0.73}Al_{0.16}Ga_{0.11}As$ QWs [18], which explains the similar characteristic temperatures $T_0$ for lasers based on structure No. 2 ($T_0$=76°C) and on four $In_{0.73}Al_{0.16}Ga_{0.11}As$ QWs ($T_0$=81°C [18]).

The external quantum efficiencies as a function of temperature for lasers with 1 mm length are presented in Figure 2, bottom panel. A change in the slope of the $\eta_{ext}(T)$ dependence is clearly observed at a temperature $T_c$. Below this temperature, the internal loss of the SCH region ($\alpha_{SCH}$) are quite small compared to the sum of the internal and the mirror losses. Thus, the $\eta_{ext}(T)$ dependence can be expressed as exponential depending on temperature [52, 53]. Above $T_c$, $\alpha_{SCH}$ increases with temperature [52, 54] leading to a significant decrease in the external quantum efficiency. Large $\alpha_{SCH}$ values are caused by an increase in carrier population in the SCH and barriers layers [56]. Moreover, hole leakage has been observed with increasing temperature in InAlGaAs–InP SCH BA lasers due to a decrease in the valence band discontinuity [57], which also effects $\eta_{ext}(T)$.

The largest characteristic temperature $T_1$, defined as $1/T_1 = d\ln(\eta_{ext})/dT$ [51] is ~100 K in the range from 20 °C and 55 °C for lasers based on structure No. 2.

## V. Conclusion

An optimization of the active region for 1300 nm lasers was presented based on InGaAs-InP superlattices. Very low internal loss (~6 cm$^{-1}$) were observed for broad-area lasers with active region based on highly strained $In_{0.74}Ga_{0.26}As/In_{0.53}Al_{0.25}Ga_{0.22}As$ superlattices. The maximum external quantum efficiency for 1 mm long lasers is found to be 37% exceeding by 20% the largest external quantum efficiency for InAlGaAs–InP SCH BA lasers of 31 % reported so far [18]).

Increasing the strain mismatch between InGaAs and InP to 1.44% makes it possible to reduce the transparency current density to approximately 500 A/cm$^2$. In addition, reducing the width of $In_{0.74}Ga_{0.26}As$ to 1 nm allows one to maximize the characteristic temperatures $T_0$ and $T_1$ to about 76 K and 100 K, respectively.

The improved gain characteristics, laser temperature stability and differential gain of the superlattice-based active region is very promising for using as an alternative to the standard strain-compensated active regions based on InAlGaAs QWs for 1300 nm VCSELs, showing larger SRH-recombination and larger threshold currents [7, 8, 9, 17]. The threshold current density of the SL-based active region is expected to drop by ~23% compared to InAlGaAs QWs. In addition, the differential gain is significantly increased (at least 33 %) compared to InAlGaAs QWs. The energy-efficiency of 1300 nm VCSELs based on both, wafer-fusion and hybrid technologies is expected to improve.

## References

[1] John Wiley & Sons, Inc. (2023, Feb. 06), Industrializing SWIR VCSELs above 1300 nm [Online]. Available: https://www.wileyindustrynews.com/en/news/industrializing-swir-vcsels-above-1300-nm (accessed on 02 August 2024).

[2] Compound Semiconductor Magazine Angel Business Comms. Ltd. (2024, Apr. 21), TriEye and Vertilas demo 1.3µm VCSEL-driven SWIR sensors [Online]. Available: https://compoundsemiconductor.net/article/119206/TriEye_and_Vertilas_demo_13%CE%BCm_VCSEL-driven_SWIR_sensors (accessed on 02 August 2024).

[3] A. D. Johnson, K. Nunna, A. Clark, A. Joel, and R. Pelzel, "Long wavelength dilute nitride VCSELs, edge emitters and detectors for 3D sensing applications," in Proc. *SPIE Photonics West 2023. OPTO - Vertical-Cavity Surface-Emitting Lasers XXVII*, San Francisco, CA, USA, 2023, Art. No. PC124390A, doi: 10.1117/12.2668691.

[4] A. Clark, K. Nunna, M. J. Furlong, and R. Pelzel, "Dilute nitride based sensing on 200mm GaAs & GaAs-Ge templates," in Proc. *SPIE Photonics West 2024. OPTO - Vertical-Cavity Surface-Emitting Lasers XXVIII*, San Francisco, CA, USA, 2024, Art. No. PC129040A, doi: 10.1117/12.3013944.

[5] Juno Publishing and Media Solutions Ltd. (2024, Apr. 08), Adtran and Vertilas unveil first ultra-low-power 100G PAM4 single-mode VCSEL technology [Online]. Available: https://www.semiconductor-today.com/news_items/2024/apr/adtran-vertilas-080424.shtml (accessed on 02 August 2024).

[6] Adtran, Inc., (2024, Mar. 24), Adtran and Vertilas answer AI demands with industry-first ultra-low-power 100G PAM4 single-mode VCSEL technology [Online]. Available: https://www.adtran.com/en/newsroom/press-releases/20240325-adtran-answers-ai-demands-with-industry-first-ultra-low-power-100g-pam4-single-mode-vcsel (accessed on 02 August 2024).

[7] Vertilas, GmbH, (2024, Mar. 25), NIR Lasers (VCSEL) for optical communications, sensing and 3D sensing [Online]. Available: https://www.vertilas.com/sites/default/files/Downloads/vertilas_inp_vcsel_com_40g_nrz_to_106g_pam4_website.pdf (accessed on 02 August 2024).

[8] A. Malacarne et al., "Optical transmitter based on a 1.3-µm VCSEL and a SiGe driver circuit for short-reach applications and beyond," *J. Lightwave Technol.*, vol. 36, no. 9, pp. 1527–1536, May 2018, doi: 10.1109/jlt.2017.2782882.

[9] L. Breyne et al., "DSP-free and real-time NRZ transmission of 50 Gb/s over 15-km SSMF and 64 Gb/s back-to-back with a 1.3-µm VCSEL," *J. Lightwave Technol.*, vol. 37, no. 1, pp. 170–177, Jan. 2019, doi: 10.1109/jlt.2018.2885619.

[10] M. Ortsiefer et al., "High speed 1.3 μm VCSELs for 12.5 Gbit/s optical interconnects," *Electron. Lett.*, vol. 44, no. 16, pз. 974–975, 2008, doi: 10.1049/el:20081591.

[11] R. Dohle et al., "New packaging technology for disruptive 1- and 2-dimensional VCSEL arrays and their electro- optical performance and applications," in Proc. *2022 IEEE 72nd Electronic Components and Technology Conference (ECTC)*, San Diego, CA, USA, 2022, doi: 10.1109/ectc51906.2022.00298.

[12] The European Photonics Industry Consortium (EPIC-Photonics). (2022, May 17), Single mode and multi mode long wavelength VCSELs [Online]. Available: https://epic-photonics.com/wp-content/uploads/2022/01/Christian-Neumeyr-Vertilas.pdf (accessed on 02 August 2024).

[13] Compound Semiconductor Magazine Angel Business Comms. Ltd. (2023, Mar. 16), Multiple opportunities for long-wavelength VCSELs [Online]. Available: https://compoundsemiconductor.net/article/116339/Multiple_opportunities_for_long-wavelength_VCSELs (accessed on 02 August 2024).

[14] P. Wolf et al., "Spectral efficiency and energy efficiency of pulse-amplitude modulation using 1.3 μm wafer-fusion VCSELs for optical interconnects," *ACS Photonics*, vol. 4, no. 8, pp. 2018–2024, Aug. 2017, doi: 10.1021/acsphotonics.7b00403.

[15] A. Caliman et al., "25 Gbps direct modulation and 10 km data transmission with 1310 nm waveband wafer fused VCSELs," *Opt. Express*, vol. 24, no. 15, pp. 16329–16335, Jul. 2016, doi: 10.1364/oe.24.016329.

[16] D. Ellafi et al., "Effect of cavity lifetime variation on the static and dynamic properties of 1.3-μm wafer-fused VCSELs," *IEEE J. Sel. Top. Quantum Electron.*, vol. 21, no. 6, pp. 414–422, Nov. 2015, doi: 10.1109/jstqe.2015.2412495.

[17] M. Müller et al., "1.3 μm high-power short-cavity VCSELs for high-speed applications," in Proc. *CLEO: Science and Innovations 2012*, San Jose, CA, USA, 2024, Art. No. CW3N.2, doi: 10.1364/cleo_si.2012.cw3n.2.

[18] P. Savolainen et al., "AlGaInAs/InP strained-layer quantum well lasers at 1.3 μm grown by solid source molecular beam epitaxy," *J. Electron. Mater.*, vol. 28, no. 8, pp. 980–985, Aug. 1999, doi: 10.1007/s11664-999-0208-6.

[19] G. Boehm et al., "InP-based VCSEL technology covering the wavelength range from 1.3 to 2.0 μm," in Proc. *International Conference on Molecular Bean Epitaxy, 2002*, San Francisco, CA, USA, 2002, doi: 10.1109/mbe.2002.1037768.

[20] R. Rodes et al., "High-speed 1550 nm VCSEL data transmission link employing 25 GBd 4-PAM modulation and hard decision forward error correction," *J. Lightwave Technol.*, vol. 31, no. 4, pp. 689–695, Feb. 2013, doi: 10.1109/jlt.2012.2224094.

[21] M. S. Moreolo et al., "Programmable VCSEL-based photonic system architecture for future agile Tb/s metro networks," *JOCN*, vol. 13, no. 2, pp. A187-A199, Dec. 2020, doi: 10.1364/jocn.411964.

[22] M. Svaluto Moreolo et al., "VCSEL-based sliceable bandwidth/bitrate variable transceivers," in Proc. *SPIE Photonics West 2019. OPTO - Metro and Data Center Optical Networks and Short-Reach Links II*, San Francisco, CA, USA, 2019, Art. No. 1094606, doi: 10.1117/12.2509316.

[23] S. Paul et al., "10-Gb/s direct modulation of widely tunable 1550-nm MEMS VCSEL," *IEEE J. Sel. Top. Quantum Electron.*, vol. 21, no. 6, pp. 436–443, Nov. 2015, doi: 10.1109/jstqe.2015.2418218.

[24] F. Riemenschneider, "Mikromechanisch abstimmbare, vertikal emittierende Laserdioden," Ph.D. dissertation, Fachbereich Elektrotechnik und Informationstechnik, Technischen Universiẗat Darmstadt, Darmstadt, Ger., 2008. [Online]. Available: https://tuprints.ulb.tu-darmstadt.de/1288/1/Mikromechanisch_abstimmbare_vertikal_emittierende_Laserdioden.pdf (accessed on 02 August 2024).

[25] M. T. Haidar et al., "Systematic characterization of a 1550 nm microelectromechanical (MEMS)-tunable vertical-cavity surface-emitting laser (VCSEL) with 7.92 THz tuning range for terahertz photomixing systems," *J. Appl. Phys.*, vol. 123, no. 2, Art. No. 023106, Jan. 2018, doi: 10.1063/1.5003147.

[26] S. Paul, P. Martín-Mateos, N. Heermeier, F. Küppers, and P. Acedo, "Multispecies heterodyne phase sensitive dispersion spectroscopy over 80 nm using a MEMS-VCSEL," *ACS Photonics*, vol. 4, no. 11, pp. 2664–2668, Oct. 2017, doi: 10.1021/acsphotonics.7b00704.

[27] A. Babichev et al., "Single-mode high-speed 1550 nm wafer fused VCSELs for narrow WDM systems," *IEEE Photonics Technol. Lett.*, vol. 35, no. 6, pp. 297–300, Mar. 2023, doi: 10.1109/lpt.2023.3241001.

[28] A. V. Babichev et al., "6-mW single-mode high-speed 1550-nm wafer-fused VCSELs for DWDM application," *IEEE J. Quantum Electron.*, vol. 53, no. 6, pp. 1–8, Dec. 2017, doi: 10.1109/jqe.2017.2752700.

[29] A. V. Babichev et al., "Continuous wave and modulation performance of 1550nm band wafer-fused VCSELs with MBE-grown InP-based active region and GaAs-based DBRs," in Proc. *SPIE Photonics West 2017. OPTO - Vertical-Cavity Surface-Emitting Lasers XXI*, San Francisco, CA, USA, 2017, Art. No. 1012208, doi: 10.1117/12.2250842.

[30] A. Babichev et al., "Impact of device topology on the performance of high-speed 1550 nm wafer-fused VCSELs," *Photonics*, vol. 10, no. 6, Art. No. 660, Jun. 2023, doi: 10.3390/photonics10060660.

[31] T. Ishikawa et al., "1.3-μm AlGaInAs/InP strained multiple quantum well lasers for high-temperature operation," Technical Digest. Summaries of papers presented at the Conference on Lasers and Electro-Optics. Conference Edition. 1998 Technical Digest Series, Vol.6 (IEEE Cat. No.98CH36178), 1998, doi: 10.1109/cleo.1998.676185.

[32] Y. Kawamura, A. Wakatsuki, Y. Noguchi, and H. Iwamura, "InGaAs/InGaAlAs MQW lasers with InGaAsP guiding layers grown by gas source molecular beam epitaxy," *IEEE Photonics Technol. Lett.*, vol. 3, no. 11, pp. 960–962, Nov. 1991, doi: 10.1109/68.97826.

[33] J. P. Loehr and J. Singh, "Theoretical studies of the effect of strain on the performance of strained quantum well lasers based on GaAs and InP technology," *IEEE J. Quantum Electron.*, vol. 27, no. 3, pp. 708–716, Mar. 1991, doi: 10.1109/3.81381.

[34] Y. Kawamura, H. Asai, Y. Sakai, I. Kotaka, and M. Naganuma, "InGaAs/InAlAs SCH-MQW lasers with superlattice optical confinement layers grown by MBE," *IEEE Photonics Technol. Lett.*, vol. 2, no. 1, pp. 1–2, Jan. 1990, doi: 10.1109/68.47022.

[35] S. A. Blokhin et al., "High power single mode 1300-nm superlattice based vcsel: impact of the buried tunnel junction diameter on performance," *IEEE J. Quantum Electron.*, vol. 58, no. 2, pp. 1–15, Apr. 2022, doi: 10.1109/jqe.2022.3141418.

[36] M.-C. Amann and W. Hofmann, "InP-based long-wavelength VCSELs and VCSEL Arrays," *IEEE J. Sel. Top. Quantum Electron.*, vol. 15, no. 3, pp. 861–868, 2009, doi: 10.1109/jstqe.2009.2013182.

[37] G. Boehm, M. Ortsiefer, R. Shau, F. Koehler, R. Meyer, and M.-C. Amann, "AlGaInAs/InP-epitaxy for long wavelength vertical-cavity surface-emitting lasers," *J. Cryst. Growth*, vol. 227–228, pp. 319–323, Jul. 2001, doi: 10.1016/s0022-0248(01)00713-8.

[38] M. Ortsiefer et al., "High speed 1.3 μm VCSELs for 12.5 Gbit/s optical interconnects," *Electron. Lett.*, vol. 44, no. 16, pp. 974–975, 2008, doi: 10.1049/el:20081591.

[39] S. A. Blokhin et al., "20-Gbps 1300-nm range wafer-fused vertical-cavity surface-emitting lasers with InGaAs/InAlGaAs superlattice-based active region," *Opt. Eng.*, vol. 61, no. 09, Art. No. 096109, Sep. 2022, doi: 10.1117/1.oe.61.9.096109.

[40] S. Blokhin et al., "Wafer-fused 1300 nm VCSELs with an active region based on superlattice," *Electron. Lett.*, vol. 57, no. 18, pp. 697–698, Jun. 2021, doi: 10.1049/ell2.12232.

[41] G. G. Zegrya, N. A. Pikhtin, G. V. Skrynnikov, S. O. Slipchenko, and I. S. Tarasov, "Threshold characteristics of λ=1.55 μm InGaAsP/InP heterolasers," *Semiconductors*, vol. 35, no. 8, pp. 962–969, Aug. 2001, doi: 10.1134/1.1393036.

[42] M. Rosenzweig, M. Mohrle, H. Duser, and H. Venghaus, "Threshold-current analysis of InGaAs-InGaAsP multiquantum well separate-confinement lasers," *IEEE J. Quantum Electron.*, vol. 27, no. 6, pp. 1804–1811, Jun. 1991, doi: 10.1109/3.90008.

[43] H. Shimizu, K. Kumada, N. Yamanaka, N. Iwai, T. Mukaihara, and A. Kasukawa, "1.3-μm InAsP modulation-doped MQW lasers," *IEEE J. Quantum Electron.*, vol. 36, no. 6, pp. 728–735, Jun. 2000, doi: 10.1109/3.845730.

[44] S. R. Selmic et al., "Design and characterization of 1.3-μm AlGaInAs-InP multiple-quantum-well lasers," *IEEE J. Sel. Top. Quantum Electron.*, vol. 7, no. 2, pp. 340–349, 2001, doi: 10.1109/2944.954148.

[45] S. Spiga, D. Schoke, A. Andrejew, G. Boehm, and M.-C. Amann, "Effect of cavity length, strain, and mesa capacitance on 1.5-μm VCSELs performance," *J. Lightwave Technol.*, vol. 35, no. 15, pp. 3130–3141, Aug. 2017, doi: 10.1109/jlt.2017.2660444.

[46] Y. Seko and A. Sakamoto, "Valence subband structures and optical properties of strain-compensated quantum wells," *Jpn. J. Appl. Phys.*, vol. 40, no. 1R, Art. No. 34, Jan. 2001, doi: 10.1143/jjap.40.34.

[47] M. P. C. M. Krijn, "Heterojunction band offsets and effective masses in III-V quaternary alloys," *Semicond. Sci. Technol.*, vol. 6, no. 1, pp. 27–31, Jan. 1991, doi: 10.1088/0268-1242/6/1/005.

[48] A. I. Akhtar and J. M. Xu, "Differential gain in coupled quantum well lasers," *J. Appl. Phys.*, vol. 78, no. 5, pp. 2962–2969, Sep. 1995, doi: 10.1063/1.360043.

[49] E. Kapon "Semiconductor Lasers I, Fundamentals," in *Optics and Photonics*, P. L. Kelley, I. P. Kaminow, G. P. Agrawal, Eds. Academic Press: San Diego, 1999, doi: 10.1016/b978-0-12-397630-7.x5000-8.

[50] J. Piprek, Y. A. Akulova, D. I. Babic, L. A. Coldren, and J. E. Bowers, "Minimum temperature sensitivity of 1.55 μm vertical-cavity lasers at −30 nm gain offset," *Appl. Phys. Lett.*, vol. 72, no. 15, pp. 1814–1816, Apr. 1998, doi: 10.1063/1.121318.

[51] N. Hossain et al., "The influence of growth conditions on carrier recombination mechanisms in 1.3 μm GaAsSb/GaAs quantum well lasers," *Appl. Phys. Lett.*, vol. 102, no. 4, Jan. 2013, doi: 10.1063/1.4789859.

[52] S. Seki, H. Oohashi, H. Sugiura, T. Hirono, and K. Yokoyama, "Study on the dominant mechanisms for the temperature sensitivity of threshold current in 1.3-μm InP-based strained-layer quantum-well lasers," *IEEE J. Quantum Electron.*, vol. 32, no. 8, pp. 1478–1486, 1996, doi: 10.1109/3.511561.

[53] M. Ohta et al., "Low threshold GaInAs quantum well lasers grown under low growth rate by solid-source MBE for 1200 nm wavelength range," *Phys. Status Solidi C*, vol. 3, no. 3, pp. 419–422, Feb. 2006, doi: 10.1002/pssc.200564112.

[54] S. Seki, H. Oohasi, H. Sugiura, T. Hirono, and K. Yokoyama, "Dominant mechanisms for the temperature sensitivity of 1.3 μm InP-based strained-layer multiple-quantum-well lasers," *Appl. Phys. Lett.*, vol. 67, no. 8, pp. 1054–1056, Aug. 1995, doi: 10.1063/1.114462.

[55] A. S. Dashkov, N. A. Kostromin, A. V. Babichev, L. I. Goray, and A. Yu. Egorov, "Simulation of the energy-band structure of superlattice of quaternary alloys of diluted nitrides," *Semiconductors*, vol. 57, no. 3, pp. 203–210, May 2023, doi: 10.21883/sc.2023.03.56237.4163

[56] A. A. Bernussi, H. Temkin, D. L. Coblentz, and R. A. Logan, "Effect of barrier recombination on the high temperature performance of quaternary multiquantum well lasers," *Appl. Phys. Lett.*, vol. 66, no. 1, pp. 67–69, Jan. 1995, doi: 10.1063/1.114185.

[57] Y. Arakawa and A. Yariv, "Theory of gain, modulation response, and spectral linewidth in AlGaAs quantum well lasers," *IEEE J. Quantum Electron.*, vol. 21, no. 10, pp. 1666–1674, Oct. 1985, doi: 10.1109/jqe.1985.1072555.

**Andrey Babichev** received the Ph.D. degree in condensed matter physics from the Ioffe Institute, St. Petersburg, Russia, in 2014. He has authored or co-authored 182 papers published in refereed journals and conference proceedings. His research interests include semiconductor heterostructures and lasers based on them. Repeated DAAD scholarship as well as Metchnikov scholarship holder. He received the Academia Europaea awards (the Russian Prizes) in 2016. He also admitted the Ioffe Prize in 2020 and 2023.

**Evgeniy Pirogov** interests include semiconductor heterostructures and its growth technology. He has authored or co-authored 59 papers published in refereed journals and conference proceedings.

**Maksim Sobolev** received the Ph.D. degree from the Institute for Analytical Instrumentation, St. Petersburg, Russia, in 2015. He has authored or co-authored 51 papers published in refereed journals and conference proceedings.

**Sergey Blokhin** received the Ph.D. degree in semiconductor physics from the Ioffe Institute, St. Petersburg, in 2006. He has authored or co-authored over 180 papers published in refereed journals and conference proceedings. His current research interests include technology and characterization of III–V semiconductor nanostructures and development of the different optoelectronic devices, including vertical-cavity surface emitting lasers, photodiodes and single photon sources.

**Yuri Shernyakov** was born in Leningrad, Union of Soviet Socialist Republics (USSR), in 1951. He graduated from the Leningrad Polytechnical Institute in 1974. He received the Candidate of Science (Ph.D.) degree in physics and mathematics from the Ioffe Institute in 1986. Since 1974, he has been working with the Ioffe Institute. He has coauthored more than 200 papers. His current research interests include edge emitting diode lasers based on InGaAs-GaAs material systems.

**Mikhail Maximov** graduated from the Leningrad Politechnical Institute in 1989. He received the Candidate of Science (Ph.D.) and D.Sc. (Habilitation) degrees in physics and mathematics from the Ioffe Physical-Technical Institute in 1995 and 2010, respectively. Since 1989, he has been working with the Ioffe Institute. He is currently at Alferov University, Saint Petersburg, Russia. He has coauthored more than 400 papers. His main research interests include development and optical studies of novel optoelectronic devices based on self-organized quantum dots, such as edge emitting and vertical cavity surface emitting lasers, microlasers, microcavity, and photonic crystals devices. In 2000, he was awarded by the Alexander von Humboldt Research Fellowship. He was also selected as the winner of the first IEEE J. Quantum Electron. Best Paper Award in 2001.

**Andrey Lutetskiy** received the Ph.D. degree in semiconductor physics from the Ioffe Institute, St. Petersburg, in 2009. His current research interests include edge emitting diode lasers based on InAlGaAs-InP material systems.

**Nikita Pikhtin** received the Ph.D. degree in semiconductor physics from the Ioffe Institute, St. Petersburg, in 2000. His current research interests include edge emitting diode lasers based on InAlGaAs-InP material systems.

**Leonid Karachinsky** received the Ph.D. degree in semiconductor physics from the Ioffe Institute, St. Petersburg, Russia, in 2004, and Dr. Sci. degree in Engineering from the ITMO University, St. Petersburg, Russia, 2021. Since 2021 he became a Professor and Deputy Director at the Institute of Advanced Data Transfer Systems, ITMO University. He has co-authored over 100 papers. His research interests include semiconductor heterostructures and lasers based on them.

**Innokenty Novikov** received the Ph.D. degree from the Ioffe Institute, Russian Academy of Sciences, St. Petersburg, in 2005. He has co-authored over 50 papers. His current research interests include development and investigation of the different types of semiconductor lasers and theoretical modeling of optical properties of semiconductor lasers.

**Anton Egorov** received the Diploma degree from the Electrical Engineering Institute in Leningrad in 1987 and the Ph.D. and Dr. Sci. degrees in 1996 and 2011, respectively. He is a Professor at ITMO University, St. Petersburg, Russia. He has authored over 410 papers published in refereed journals and conference proceedings. His area of experience includes molecular beam epitaxy; III–V semiconductor heterostructures; GaAs- and InP-based (In,Ga,Al) (As,N) quantum well and quantum dots heterostructures; edge- and vertical-cavity surface emitting lasers (infrared); and III–V heterostructures for microelectronics.

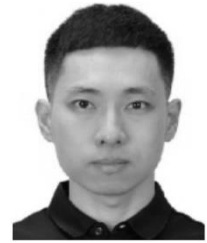

**Si-Cong Tian** received his B. Sc. and Ph. D. degrees in Physics from Jilin University, China, in 2007 and 2012, respectively. From 2016-2017 he studied at Arkansas University, US, as a visiting scholar. Currently, he is a Professor at Bimberg Chinese-German Center for Green Photonics, Changchun Institute of Optics Fine Mechanics and Physics, Chinese Academy of Sciences, China. His research interest includes high-speed vertical-cavity surface-emitting lasers and high-brightness semiconductor lasers.

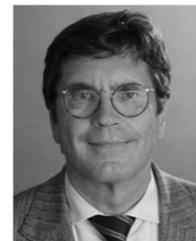

**Dieter Bimberg** (LF'16) received the Ph.D. degree from Goethe University, Frankfurt, Germany, in 1971. He held a Principal Scientist position at the Max Planck Institute for Solid State Research, Grenoble, France, until 1979. After serving as a Professor of electrical engineering with the Technical University of Aachen, Aachen, Germany, he assumed the Chair of Applied Solid-State Physics with the Technical University of Berlin, Berlin, Germany. He is the Founding Director of Center of Nanophotonics at Technical University of Berlin. He held guest professorships at the Technion, Haifa, Israel, the University of California at Santa Barbara, Santa Barbara, CA, USA, and Hewlett-Packard, Palo Alto, CA, USA. He was a Distinguished Adjunct Professor with King Abdulaziz University, Jeddah, Saudi Arabia, from 2012 to 2018, and the Einstein Professor with the Chinese Academy of Sciences in 2017 and 2018. Since 2018, he has been the Director of the Bimberg Chinese–German Center for Green Photonics, Chinese Academy of Sciences, Changchun Institute of Optics, Fine Mechanics, and Physics, Changchun, China. He has authored more than 1500 papers, 60 patents, and six books. The numbers of times his research works have been cited exceeds 58 000 and his Hirsch index is 107.

His research interests include physics and technology of nanostructures, nanostructured photonic and electronic devices and energy efficient data communication. He is a member of the Russian Academy of Sciences and the German Academy of Sciences Leopoldina, a Foreign Member of the National Academy of Engineering of USA, and a Fellow of the U.S. National Institute of Inventors. The American Physical Society and IEEE elected him as Fellow/Life Fellow, respectively. Numerous international awards, such as the UNESCO award and medal for nanoscience and technology and the Russian State Prize for Science and Technology were bestowed to him.

His research interests include physics and technology of nanostructures, nanostructured photonic and electronic devices and energy efficient data communication. He is a member of the Russian Academy of Sciences and the German Academy of Sciences Leopoldina, a Foreign Member of the National Academy of Engineering of USA, and a Fellow of the U.S. National Institute of Inventors. The American Physical Society and IEEE elected him as Fellow/Life Fellow, respectively. Numerous international awards, such as the UNESCO award and medal for nanoscience and technology and the Russian State Prize for Science and Technology were bestowed to him.